# Lightweight Joint Simulation of Vehicular Mobility and Communication with LIMoSim

Benjamin Sliwa[1], Johannes Pillmann[1], Fabian Eckermann[1], Lars Habel[2],
Michael Schreckenberg[2] and Christian Wietfeld[1]
[1]Communication Networks Institute, TU Dortmund University, 44227 Dortmund, Germany
e-mail: {Benjamin.Sliwa, Johannes.Pillmann, Fabian.Eckermann, Christian.Wietfeld}@tu-dortmund.de
[2]Physics of Transport and Traffic, University Duisburg-Essen, Germany
e-mail: {Habel, Schreckenberg}@ptt.uni-due.de

*Abstract*—The provision of reliable and efficient communication is a key requirement for the deployment of autonomous cars as well as for future Intelligent Transportation Systems (ITSs) in smart cities. Novel communications technologies will have to face highly-complex and extremely dynamic network topologies in a Vehicle-to-Everything (V2X)-context and will require the consideration of mobility information into decision processes for routing, handover and resource allocation. Consequently, researches and developers require simulation tools that are capable of providing realistic representations for both components as well as means for leveraging the convergence of mobility and communication. In this paper, we present a lightweight framework for the simulation of vehicular mobility, which has a communications-oriented perspective by design and is intended to be used in combination with a network simulator. In contrast to existing approaches, it works without requiring Interprocess Communication (IPC) using an integrated approach and is therefore able to reduce the complexity of simulation setups significantly. Since mobility and communication share the same codebase, it is able to model scenarios with a high level of interdependency between those two components. In a proof-of-concept study, we evaluate the proposed simulator in different example scenarios in an Long Term Evolution (LTE)-context using real-world map data.

## I. INTRODUCTION

With the transition from human-driven cars to autonomous traffic systems, the importance of modern communication technologies for modern traffic systems grows dramatically. Furthermore, novel communication systems will have to provide reliable and efficient means of communication for highly-mobile entities that operate in constantly changing topologies and compete about the limited resources with a large number of other network participants. A promising method for addressing these challenges is the joint consideration of mobility and communication in order to exploit their interdependencies for developing predictive and context-aware communication systems. In earlier work [1], we showed that the reliability and efficiency of data transmissions in Mobile Ad-hoc Networks (MANETs) can be highly increased by integrating knowledge about the mobility characteristics of the mobile nodes into the decision processes. In a smart city context, *mobility-aware* communication principles will be used for mobility-predictive LTE-handovers and resource allocation strategies for autonomous cars [2], which system-immanently have information about their future trajectory. Furthermore, *communication-aware* mobility is already an important research topic [3] in the context of Unmanned Aerial Vehicle (UAV) networks and could get an increased attention for fleet-monitoring in intelligent logistic systems. As mobility and communication converge (cf. Fig. 1), researchers and developers require efficient simulation tools that are capable of providing valid models of both worlds as well as possibilities for exploiting their interdependencies. Current state-of-the-art approaches for simulating vehicular communication networks rely on IPC-based coupling of multiple specialized simulators, which were developed for isolated fields of application. As a result, simulation setups have a high complexity and require the synchronization of the simulators at runtime. Since the actual algorithms for the vehicular motion and the communication behavior are treated separately, it is difficult to implement mobility-aware models. In contrast to the existing approaches, the proposed Lightweight ICT-centric Mobility Simulation (LIMoSim) was developed with an communications-oriented view on mobility by design and uses an integrative approach instead of a coupling mechanism with an IPC-bottleneck. Consequently, it is able to offer a simpler setup for simulation scenarios. Furthermore, since communication algorithms can directly access mobility information and vice versa through a shared codebase, sophisticated algorithms can be developed for exploiting the synergies between those components. The paper is structured as follows: in the next section, an overview about existing approaches for simulating vehicular networks is provided. Afterwards, we present our proposed framework and give detailed descriptions of its individual components. In the next section, we present the setup of a reference scenario for performance evaluation using Objective Modular Network Testbed in C++ (OMNeT++) [4]. Finally, detailed simulation results are presented and discussed.

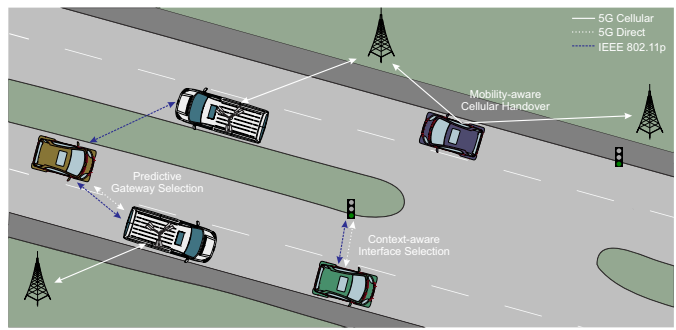

Fig. 1. The simulation of future ITS will require the joint consideration of vehicular mobility and communication. Moreover, modern communication technologies will leverage mobility information for optimizing decision processes like handovers, resource allocation, forwarder selection and predictive alignment of pencil-beams in upcoming 5G networks.

TABLE I
STATE-OF-THE-ART SIMULATORS FOR VEHICULAR MICRO MOBILITY

| Simulator | Open Source | Scalability | Focus | Comment |
|---|---|---|---|---|
| SUMO [6] | yes | high | Comparison of different mobility models | Optional external control through TraCI |
| PTV Vissim [7] | no | very high | Multi-modal traffic | Offers highly detailed scenario modeling |
| OLSIMv4 [8] | no | high | Highway-traffic | Online-coupling with real-world highway data |
| MATSim-T [9] | yes | very high | Large-scale scenarios | - |

## II. RELATED WORK

In this section, we present and discuss state-of-the-art approaches for simulating vehicular mobility and communication as well as coupling frameworks aiming to bring both aspects together. Since our research is focusing on optimizing communication processes, we consider smart city road traffic in a communications-oriented way [5], where vehicular mobility is considered a service for the simulation of communication systems.

### A. Simulation of Vehicular Mobility

Traffic simulators are commonly classified depending on their simulation granularity level. In this context, *micro mobility* describes the behavior of individual entities, whereas *macro mobility* is used for the behavior of groups as a whole. In this paper, we focus on micro mobility and agent-based models, since the simulation of communication processes requires the consideration of individual communicating entities as well.

In the following, an overview about state-of-the-art mobility simulators is presented. Tab. I summarizes the key properties of the discussed simulators.

*1) SUMO:* Simulation of Urban Mobility (SUMO) is one of the most popular traffic simulators and available as Open Source software. Its simulation kernel contains a high number of different validated models for the individual layers of vehicular mobility simulation. SUMO itself is rather a package of multiple applications than a standalone simulator, consisting of different tools for traffic simulation, visualization, data import and editing as well as preprocessing and different traffic routers. Furthermore, it supports external control through the dedicated TraCI protocol that is used by many coupling frameworks (discussed in Sec. II-B). Although it generally supports dynamic traffic scenarios using TraCI or the *Duarouter* module, its main focus are static traffic scenarios using precomputed routes and traffic flow information. While this approach is suitable for evaluating traffic phenomenons (e.g. jams) in isolated scenarios, it does not match well with the requirements of future smart city applications [10], where dynamic traffic routing through intelligent infrastructure and cooperative decision making play important roles for jam avoidance.

*2) PTV Vissim:* PTV Vissim [7] is a well-established traffic simulator with a focus on multi-modal traffic simulation with models for cars, public transport, bicycles and pedestrians and offers 2D and 3D dimensional graphical representations of the scenarios. It is able to model scenarios with very high level of details, including buildings and vegetation. The program is in a mature state and rather a commercial product for modeling traffic management systems than a framework for research projects as it is not Open Source.

*3) OLSIMv4:* OLSIMv4 [8] is focusing on the simulation of highway traffic in the German state North Rhine-Westphalia. What makes it unique and worth being mentioned despite its restricted focus, is its online-coupling with real-world loop detectors. Traffic flow measurements are brought into the simulation every minute. The data is used for short term and long term traffic forecasts.

*4) MATSim-T:* Multi Agent Transport Simulation Toolkit (MATSim-T) [9] is another agent-based traffic simulator that is well-established and focuses on large-scale scenarios with private cars and public transport. In this context, it is used for simulating big city areas and country-wide road networks.

### B. Coupled Simulation of Vehicular Mobility and Communication

Most state-of-the-art traffic simulators have been developed with an isolated focus on traffic dynamics. Many of them have their origins in traffic management systems and were designed in a time, where ITS was not yet a big research field. However, modern ITS scenarios require the integration and consideration of communication into traffic scenarios. For this purpose, coupling extensions to the state-of-the-art traffic simulators have been developed. Tab. II gives an overview about popular frameworks.

While the simplest form of coupling is the usage of *mobility traces* in communication network simulation scenarios, it only provides unidirectional interaction possibilities between mobility and communication. Since this approach cannot be used for influencing traffic routing decisions by transmitted information, only bidirectional coupling approaches are considered in the following.

*1) Veins:* Vehicles in Network Simulation (Veins) [11] is a well-established framework for OMNeT++ adding support

TABLE II
STATE-OF-THE-ART FRAMEWORKS FOR SIMULATION OF VEHICULAR COMMUNICATION NETWORKS

| Framework | Open Source | Mobility Simulator | Network Simulator | Technology | Comment |
|---|---|---|---|---|---|
| Veins [11] | yes | SUMO | OMNeT++ | IEEE 802.11p | Optional modules for platooning, channel modeling and security |
| iTETRIS [12] | yes | SUMO | NS3 | IEEE 802.11p | |
| VSimRTI [13] | no | SUMO, VISSIM | OMNeT++, NS3 | - | HLA-based coupling framework |

for the IEEE 802.11p and IEEE 1609.4 Dedicated Short-range Communications (DSRC) / Wireless Access in Vehicular Environments (WAVE) network layers. Furthermore, it acts as an interface for the vehicular traffic simulator SUMO [6] with Transmission Control Protocol (TCP)-based IPC using the TraCI protocol. Veins itself has been subject for various extensions focusing on specialized topics such as security, platooning and channel modeling.

*2) iTETRIS:* Similar to Veins, iTETRIS attaches to SUMO for the simulation of vehicular motion. The network simulation is performed with NS3 and focuses on large-scale simulations using the dedicated iTETRIS Network simulator Control Interface (iNCI).

*3) VSimRTI:* V2X Simulation Runtime Infrastructure (VSimRTI) is an High Level Architecture (HLA)-inspired coupling framework for simulators. Mobility simulations can be performed with SUMO or PTV Vissim, OMNeT++ and NS3 are supported for the network simulation among others. Furthermore, different application-layer simulators for environmental considerations can be embedded into the setup. In contrast to the other considered coupling frameworks, VSimRTI is not published as Open Source.

*Conclusion:* A general observation is that most of the frameworks are missing a clear focus. They do not only provide an implementation for a specified communication stack but also include a coupling mechanism to a road traffic simulator. Consequently, the vehicular mobility simulation is tied to a specific communication technology. Since all of the discussed network simulators support a modular modeling approach, an improved reusability of the code could be achieved by a stricter separation of mobility and communication technology. With coupling the mobility behavior to the network simulator itself instead of a specific communication technology, it could be used in different fields of application (e.g. LTE and MANETs scenarios). This is especially important if different types of vehicles (eg. cars, trains, boats, UAVs) and heterogeneous communication technologies should be evaluated together in a single simulation scenario. Furthermore, the IPC-based coupling as shown in Fig. 2 used by the state-of-the-art approaches is another bottleneck for simulative evaluations and the development process in general. Although it is able to achieve highly reliable results by using a combination of multiple specialized simulators, it has a number of disadvantages:

- *High simulation setup complexity* since multiple tools needed to be run in parallel. The effect is even more drastic for massive evaluation setups using multiple servers in parallel.
- *Compatibility drifts* occur since the simulators are developed individually without considering their interdependencies.
- The considered traffic simulators were developed for mostly static evaluation scenarios using precomputed routes, which does not match well with the specified ITS usecase. Moreover, in that field of application the computation effort for the mobility is negligible compared to the communication simulation as shown in Sec. IV-A.
- Since the data exchange between the simulators is bound to a dedicated protocol, it is difficult to develop algorithms that exploit the synergies of mobility and communication and operate on both aspects simultaneously.

Future applications and technologies for ITSs in smart cities will require the joint consideration of mobility and communication on the mobility side as well as on the communication side. While this challenge is currently being addressed by coupling multiple specialized simulators, we think that simulation tools in this area should consider mobility and communication as well as their interdependencies by design and hereby advocate for an integrated approach with a common codebase and a focus on highly dynamic traffic scenarios.

III. DESIGN AND IMPLEMENTATION OF THE SIMULATOR

The proposed simulator LIMoSim focuses on highly dynamic traffic scenarios where all decision processes and routes are determined at runtime. Its main field of application is the simulation of medium-sized city scenarios, where intelligent vehicles interact with other traffic participants through means of communication. Therefore, it provides a deep integration with network simulators using a common codebase without tying the mobility simulation to a specific communication technology. The source code is published in an Open Source way in order to enable own extensions by the user. LIMoSim is composed of two main submodules: the *simulation kernel* and the *user inferface* that can either be used in combination or independently. Fig. 3 provides an illustration of the logical system model. For ITS applications, the *managed* mode is used, where an external network simulator takes control about the event scheduling and communication simulation and only the simulation kernel of LIMoSim is used for the mobility handling. Because of the integrated coupling approach, objects from the network simulator can directly interact with LIMoSim objects and exchange information. In order to ensure wide compatibility with network simulators, the kernel consists of plain C++ code only and does not contain external dependencies. Additionally, the *standalone* mode is intended for mobility-only simulations as well as for setting up new simulation scenarios with the integrated editor. The user interface is written with Qt-C++. In the following subchapters, detailed descriptions for the individual components are provided.

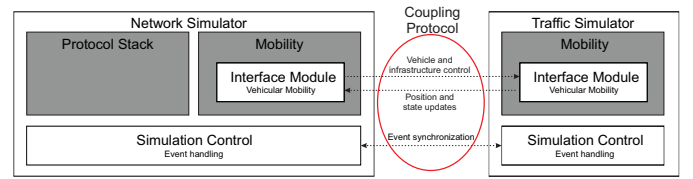

Fig. 2. IPC-based coupling of mobility and communication simulators. The IPC approach requires a protocol for the synchronization and the exchange of data between multiple specialized simulators.

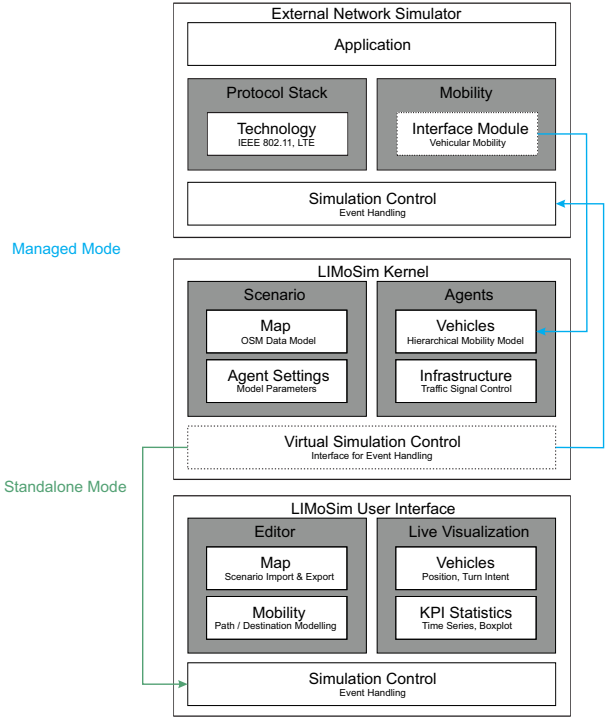

Fig. 3. Components of the proposed LIMoSim. The simulator can either operate in *managed* mode, where an external network simulator takes control about the event handling mechanisms or in *standalone* mode using an optional user inferface. As opposed to the IPC-based approach, the coupling is realized by direct code references and does not require a dedicated protocol.

### A. Agent-based Mobility Modeling

Since LIMoSim follows communications-oriented principles, it relies on selected well-known mobility models rather than having a wide range of different models. As [14] points out, the state-of-the-art microscopic simulation models behave very similar in the considered ITS application scenario anyway. Fig. 4 shows the hierarchical model, which can be mapped to the *strategic, tactical* and *operational* layers of the Hoogendorn model [15]. On the highest layer, a *strategic mobility* model is used to integrate the occasion for the vehicle's movement and to the determine the destination of the trip. The planned trajectory consisting of a list of nodes is then obtained using the routing module. The nodes are sequentially approached by the general mobility handling of the vehicle, which handles the movement on the current lane with respect to the traffic rules and priority handling. For lane changes and overtaking, the Minimizing Overall Braking Induced by Lane change (MOBIL) [16] model is applied. *Adaptive cruise control* is performed by a following model, which steers acceleration and braking depending on the other traffic participants. LIMoSim uses the well-known Intelligent Driver Model (IDM) [17] for this task. It should be noted that since the regular version of this model is not intersection-aware, the IDM implementation of LIMoSim treats traffic signals like static vehicles if the light is yellow or red. Finally, the acceleration increment is used to update the velocity and the position of the vehicle for the current simulation step.

### B. Data model and Road Editor

LIMoSim uses the components of the OpenStreetMap (OSM) data model for managing the geographical data, which consists of three basic primitives. *Nodes* are the most atomic elements and assign a logical meaning to a position value. They are used for describing street vertices as well as characteristic places like stores, parking spaces and traffic signals. Node lists consisting of at least two elements form *ways*, which are used to describe road sections as well as the shapes of buildings. In the context of road sections, a way describes a part of a road with equal properties (e.g. number of lanes, speed limitation). Consequently, most roads consist of a list of ways and are not necessarily described by a single way. Two ways intersect if they have at least one shared node. Intersections are generated automatically or can be defined

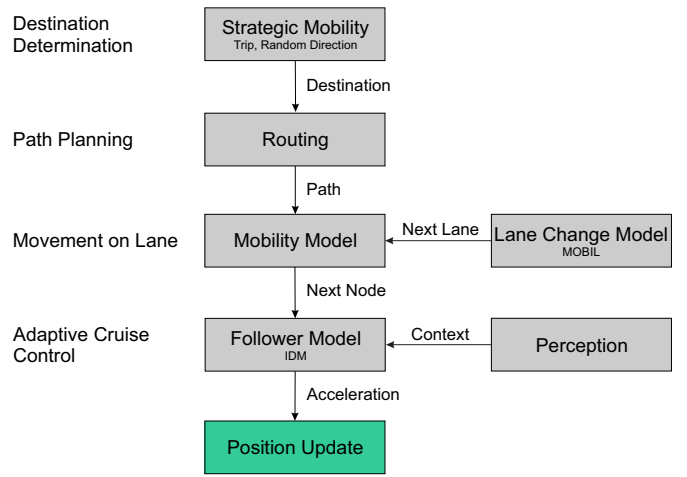

IDM: Intelligent Driver Model
MOBIL: Minimizing Overall Braking Induced by Lane change

Fig. 4. Architecture of the hierarchical mobility model.

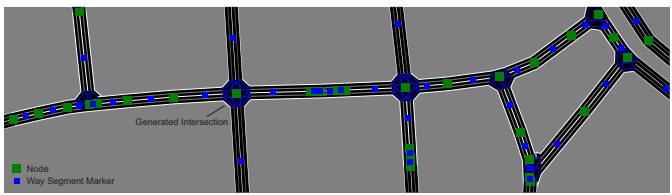

Fig. 5. Editor view for an example map excerpt. Streets consists of way segments and nodes that can be moved, added and removed

manually. *Relations* assign logical meanings to a set of ways and nodes in a given region, e.g. streets consisting of multiple road sections and houses.

OSM data can directly be imported into simulation scenarios. Alternatively, street topologies can be created and edited with the integrated road editor or using generic models (e.g. manhattan grids). Fig. 5 shows an example map excerpt which is being edited. Alternatively, established OSM-editors like Java-OpenStreetMap-Editor (JOSM) can be used. In contrast to other simulators, LIMoSim does not require preprocessing with external tools in order to load map data. When an OSM file map.osm is loaded for the first time, LIMoSim creates an optimized file map.osm.limo, which removes redundant information and contains coordinate transformations for faster network initialization when the map is accessed again.

The graphical map representation can be exported in a vectorized format, e.g. for being used for illustrating the simulation setup in scientific publications (cf. Fig. 7 as an example).

### C. Live Visualization and Export of Statistical Information

The development of novel mobility algorithms is a complex process which usually involves numerous verification steps. Many behavioral effects cannot be analyzed by considering the visual vehicular motion only and require the analysis of mobility traces. Usually this is done offline after preprocessing of log files. In order to speed this process up, LIMoSim features live visualization of relevant performance indicators for individual or multiple vehicles combined which be either displayed in the time-domain or as a boxplot. With this approach plausibility checks can be performed at runtime. The Figures can be exported in a vectorized format. Fig. 6 shows an example plot of the speed dynamics for multiple vehicles. The figure has been exported directly from LIMoSim.

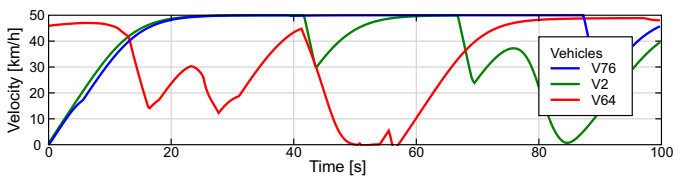

Fig. 6. Example velocity behavior of multiple vehicles in the time domain as shown by the live visualization. The plot has been exported by LIMoSim without additional graphical processing.

### D. Integration into Network Simulators

In contrast to IPC-based coupling approaches, there is no need for event-synchronization with the coupled network simulator that would add another layer of complexity to the system. Instead, the events of LIMoSim are directly embedded into the event queue of the network simulator and handled by the respective event scheduler. Since LIMoSim is provided in an Open Source way, it can be integrated into further network simulators with acceptable effort. For a detailed description about how the proposed framework is embedded into OMNeT++, cf. [18]. Moreover, the available implementation of the OMNeT++-interface can be considered a blueprint for further integration attempts.

## IV. PROOF OF CONCEPT EVALUATION

In this section, we present the setup and the results of the proof of concept evaluation. The map for the reference scenario in the city area of Dortmund is shown in Fig. 7 and consists of different types of streets with intersections and traffic signals. The driver behavior is modeled using a velocity factor $\hat{v}$, which is multiplied with the allowed velocity $v_{limit}$ in order to mirror the drivers tendency of driving slower or faster than a given limit.

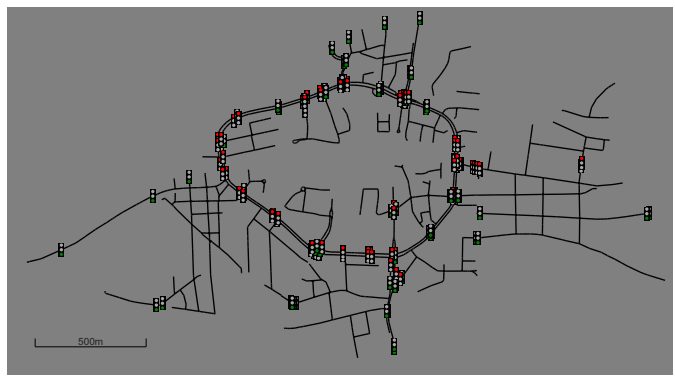

Fig. 7. Reference scenario in the city area of Dortmund using OSM location data. The figure has been exported directly from the UI part of LIMoSim using its vector graphic export function.

The default parameters for the performance evaluation are defined in Tab. III.

### A. Performance of the Mobility Simulation

Since LIMoSim is using a dynamic approach for traffic control and does not rely on precomputed routing files like state-of-art simulators, the impact of the mobility component on the overall simulation time is being analyzed. Three different evaluation scenarios are defined as follows:

- *Pure Vehicular Mobility* does not simulate any communication processes at all and is used to determine the required computation resources for solely mobility simulation.
- *LTE-based Voice over IP (VoIP)* is based on the test_handover example scenario provided by SimuLTE. Multiple vehicles move in the specified city area, causing handovers at runtime.

TABLE III
SIMULATION PARAMETERS FOR THE REFERENCE SCENARIO

| Mobility System | |
|---|---|
| Parameter | Value |
| Following model | IDM |
| Free acceleration exponent $\gamma$ | 4 |
| Desired time gap $T$ | 1 m |
| Jam distance | 2.0 m |
| Maximum acceleration | 1.4 m/s$^2$ |
| Desired deceleration | 2.0 m/s$^2$ |
| Velocity factor $\hat{v}$ | $1 \pm 0.2$ |
| Lane change model | MOBIL |

| LTE Communication System | |
|---|---|
| Parameter | Value |
| Carrier frequency | 1800 Mhz |
| eNode B transmission power | 46 dBm |
| eNode B antenna | Omnidirectional |

| MTC Application | |
|---|---|
| Parameter | Value |
| Sensor frequency | 1 Hz |
| Sensor payload | 20 kByte |
| Transmission interval $\Delta t$ | 30 s |
| Transport Layer protocol | TCP |

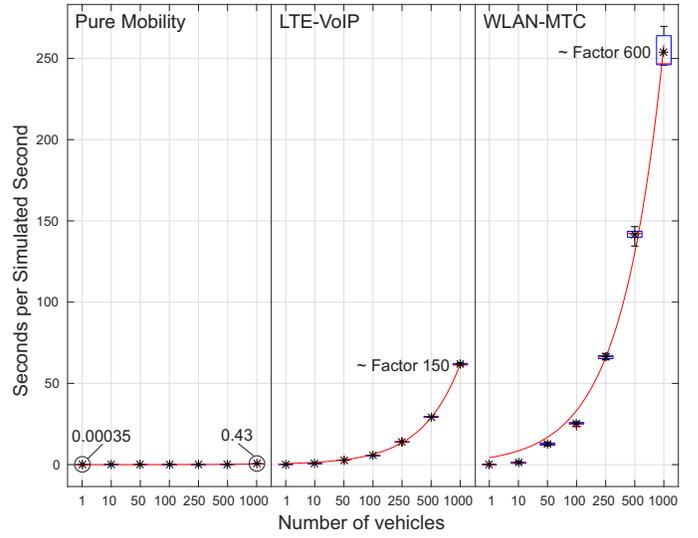

Fig. 8. Influence of mobility and communication on the total execution time of the simulation using three different evaluation scenarios in OMNeT++.

- *WLAN-based Machine-Type Communication (MTC)* simulates a number of vehicles gathering sensor data for a *crowdsensing-based* MTC application. The sensor generates 20 kByte of data per second, which is transmitted to a server using roadside units as gateways if available. The transmission is based on TCP and triggered every 30 seconds.

The simulations were performed on a Intel Core i7-4770 @ 3.40 GHz (4 cores) with 32 GB memory and Ubuntu 16.04 LTS operating systems. For fairness reasons, all simulations were performed using the OMNeT++ execution environment, even if pure mobility was simulated without any communication. All evaluations were performed 10 times with different random seeds and for the simulated time of 300 s. Fig. 8 shows the results of the performance evaluation. It is highly significant, that the simulation of the communication processes is much more complex and resource-consuming than the mobility simulation. Furthermore, the impact of the mobility component on the overall computation time can be considered negligible for most scenarios. Consequently, for ITS scenarios with a typical amount of communication, there is no need for saving computation time by using precomputed mobility data.

### B. LTE-scenario using Real-world Map Data

For the identification of general characteristics, the temporal behavior of the trip of a single example car is considered. The car is moving on the city-ring of Dortmund (cf. Fig. 7) and passes the coverage areas of multiple LTE Evolved Node Bs (eNBs). Fig. 9 shows the runtime behavior of the vehicle's velocity and acceleration as well as the RSSI to the current eNB and LTE-handover occurrences. Because of the urban environment with traffic signals and interactions with other traffic participants, velocity and acceleration show a very dynamic behavior with alternating periods of free-flow and stop-and-go traffic. Due to the mobility behavior of the vehicle, multiple LTE-handovers occur during the simulation run. For comparison, cf. [19], where a similar evaluation was performed using a different methodical setup with SUMO and an analytical LTE model, which achieved similar results for the channel characteristics and the handover behavior.

### C. Mobility-aware Example Application: Channel-aware Transmissions with Crowdsensing-based Network Quality Maps

Since one of the unique features of LIMoSim is the joint consideration of mobility and communication, an example application that exploits the synergies of both components is considered. From the total amount of vehicles of the reference scenario, a subset of 100 cars is chosen randomly to act as mobile sensors, that transmit MTC data to a cloud server. The other vehicles act as interference traffic. As the mobile sensors move in the considered area, they monitor the Signal-to-Noise Ratio (SNR) of the LTE signal and use their measurements to update an SNR map $M$ that consists of cells with cellsize $c$. For the actual data transmission, two schemes are compared. The *periodic* approach uses a fixed interval $\Delta t$ for the transmission timing. In contrast to that, the *predictive* scheme uses mobility prediction and channel-quality estimation to find the best transmission time in the available time interval centered around $\Delta t$. The general approach is to transmit data early if the channel quality is decreasing in order to save resources and to avoid retransmissions. Correspondingly, transmissions are delayed if the channel quality is improving. Due to the integrated approach with a shared codebase, knowledge about the planned trajectories of the vehicles is available and can be leveraged by the mobile sensors to optimize their communication process. After a transmission at time $t_{tx}$, the next transmission time $t'_{tx}$ is chosen with Eq. 1 as the time with the best predicted SNR for all integer values of $\tilde{t}$ in the considered interval:

$$t'_{tx} = \max_{S\tilde{N}R} \left( \tilde{t} \in \left[ t_{tx} + \Delta t - \frac{\Delta t}{2}, t_{tx} + \Delta t + \frac{\Delta t}{2} \right] \right) \quad (1)$$

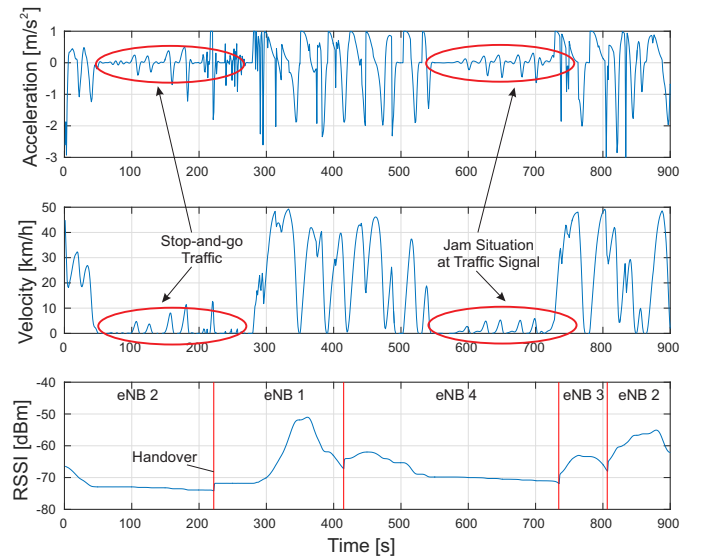

Fig. 9. Example temporal behavior characteristics for velocity, acceleration and RSSI of an LTE signal in an urban environment.

It is assumed that the vehicles have offline-access to the network quality map $M$. For the simulative evaluation, this was achieved through performing a training-run with a different random seed. The estimated SNR can be obtained from $M$ as $S\tilde{N}R = \bar{M}(C)$ for a given cell $C$. Therefore, the future position of the vehicle needs to be predicted and mapped to its cell value. With the available trajectory information, the future position $\vec{P}(t+\tau)$ with $t+\tau = \tilde{t}$ is estimated using information about the currently approached node $N$, the current vehicle position $\vec{P}_i$ and the velocity $v$.

$$\vec{P}(t+\tau) = \vec{P}(t) + \frac{\vec{N} - \vec{P}(t)}{||\vec{N} - \vec{P}(t)||} \cdot \tau \cdot v \quad (2)$$

The cell value of the predicted position can be obtained as $C(t+\tau) = \left\lfloor \frac{\vec{P}(t+\tau)}{c} \right\rfloor$. Fig. 10 shows the goodput and the transmission duration for both compared transmission schemes. The predictive approaches achieves a significantly higher goodput than the periodic scheme. Furthermore, it is able to reduce the mean transmission time by 40% and reduces the probability for extremely long transmissions. Both aspects have a positive impact in coexistence-scenarios, where MTC is competing with Human-to-Human (H2H) communication about the available radio resources [20]. Although the considered mobility prediction scheme is simple and assumes constant velocities, the results show that the joint consideration of mobility and communication is a promising method for improving the efficiency of mobile communication systems.

### D. Traffic Flow Dynamics

Finally, for the verification of the IDM implementation, a typical inner-city scenario is considered. A number of vehicles is approaching a traffic signal that is initially switching from yellow light to red and turns to green after a period of time. Fig. 11 shows a space-time diagram of the cars' trajectories and illustrates the traffic flow dynamics of the IDM following model. Different driver types can be identified according to the gaps between the different graphs. The impact of a slow driver on the following cars can be clearly seen. After the traffic signal turns green, the cars accelerate similarly, but after some time has passed the different velocity behaviors can be identified again with the slow driver slowing down his followers again. The simulated behavior characteristics of LIMoSim match well with the comprehensive analytical and simulative evaluations of the IDM that have been performed by Treiber in [21].

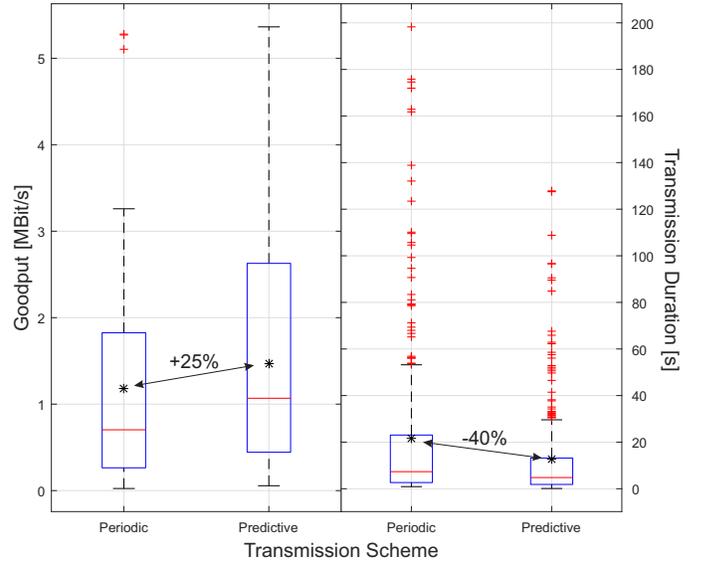

Fig. 10. Exploiting mobility knowledge for efficient communication: Comparison of goodput and transmission duration for periodic and predictive transmission schemes.

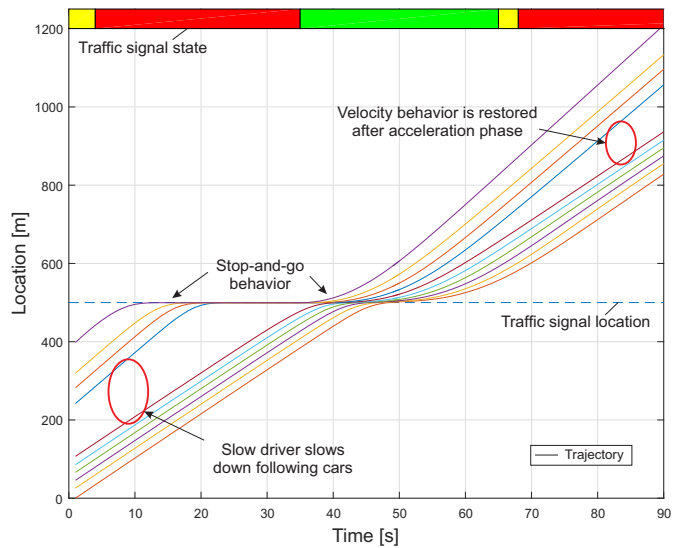

Fig. 11. Characteristics of the IDM car following model for 10 vehicles in inner city traffic influenced by a traffic signal.

## V. Conclusion and Future Work

In this paper, we have presented LIMoSim[1] as a novel Open Source framework for simulating urban vehicular traffic with support for real-world map data from OpenStreetMap (OSM). The simulation of vehicular motion relies on well-known analytical models in order to ensure precise and comparable results. In contrast to existing approaches, the kernel of the mobility simulation is meant to be embedded into the code of network simulators like OMNeT++, eliminating the need for IPC as a coupling mechanism for different simulators. Consequently, the complexity of simulation setups is highly reduced and the interoperability of mobility and communication is enhanced since both worlds share a common codebase. The capabilities of LIMoSim to provide a realistic representation of vehicular mobility for ITS applications in a lightweight and efficient way were shown with a proof of concept evaluation. Moreover, the general advantages of the joint consideration of mobility behavior and communication were demonstrated.

Currently, we are working on an integration for the OMNeT++ Vehicular Ad-hoc Network (VANET) framework Veins for having LIMoSim as an IPC-free alternative to SUMO in an IEEE 802.11p context. In future work, we want to provide support for further communication simulation frameworks, currently an integration for NS3 is being developed. Additionally, we want to emphasize the impact of human-decision making and integrate more realistic strategic mobility models (e.g. Small Worlds In Motion (SWIM) [22]) into the simulator. As the application range is not limited to urban vehicular scenarios, we are also planning to extend the framework for the multi-modal simulation of mobile robotic networks in a indoor logistic context.

### Acknowledgment

Part of the work on this paper has been supported by Deutsche Forschungsgemeinschaft (DFG) within the Collaborative Research Center SFB 876 "Providing Information by Resource-Constrained Analysis", project B4 and "European Regional Development Fund" (EFRE) 2014-2020 in the course of the "InVerSiV"project under grant number EFRE-0800422 and has been conducted within the AutoMat (Automotive Big Data Marketplace for Innovative Cross-sectorial Vehicle Data Services) project, which received funding from the European Union's Horizon 2020 (H2020) research and innovation programme under the Grant Agreement No 644657.

---

[1] Available at https://github.com/BenSliwa/LIMoSim